# The Case for Probe-class NASA Astrophysics Missions

Lead Author: **Martin Elvis**; email: melvis@cfa.harvard.edu; phone: 617 495 7442
Thematic Activity: Space Based Activity

Authors:

**Jon Arenberg** (Northrop Grumman) <jon.arenberg@ngc.com>,
**David Ballantyne** (Georgia IT) <david.ballantyne@physics.gatech.edu>,
**Mark Bautz** (MIT) <mwb@space.mit.edu>,
**Charles Beichman** (JPL) <chas@ipac.caltech.edu>,
**Jeffrey Booth** (JPL) <jeffrey.t.booth@jpl.nasa.gov>,
**James Buckley** (Washington U., St. Louis) <buckley@physics.wustl.edu>,
**Jack O. Burns** (U.Colorado, Boulder) <Jack.Burns@colorado.edu>,
**Jordan Camp** (NASA GSFC) <jordan.b.camp@nasa.gov>,
**Alberto Conti** (Ball Aerospace) <aconti@ball.com>,
**Asantha Cooray** (UC Irvine) <acooray@uci.edu>,
**William Danchi** (NASA GSFC) <william.c.danchi@nasa.gov>,
**Jacques Delabrouille** (CNRS/APC Paris, CEA Saclay) <delabrou@apc.in2p3.fr>,
**Gianfranco De Zotti** (INAF) <gianfranco.dezotti@inaf.it>,
**Raphael Flauger** (UC San Diego) <flauger@physics.ucsd.edu>,
**Jason Glenn** (U.Colorado, Boulder) <jason.glenn@colorado.edu>,
**Jonathan Grindlay** (Harvard/CfA) <jgrindlay@cfa.harvard.edu>,
**Shaul Hanany** (U. MN) <hanany@umn.edu>,
**Dieter Hartmann** (Clemson) <hdieter@g.clemson.edu>,
**George Helou** (IPAC) <gxh@ipac.caltech.edu>,
**Diego Herranz** (CSIC-UC, Santander) <herranz@ifca.unican.es>,
**Johannes Hubmayr** (NIST) <johanneshubmayr@gmail.com>,
**Bradley R. Johnson** (Columbia) <bradley.johnson@columbia.edu>,
**William Jones** (Princeton) <wcjones@princeton.edu>,
**N. Jeremy Kasdin** (Princeton) <jkasdin@princeton.edu>,
**Chryssa Kouveliotou** (G.Washington U.) <ckouveliotou@email.gwu.edu>,
**Kerstin E. Kunze** (U. Salamanca) <kkunze@usal.es>,
**Charles Lawrence** (JPL) <charles.lawrence@jpl.nasa.gov>,
**Joseph Lazio** (JPL) joseph.lazio@jpl.nasa.gov,
**Sarah Lipscy** (Ball Aerospace) slipscy@ball.com,
**Charles F. Lillie** (Lillie Consulting LLC) <charles.lillie@clillie.com>
**Tom Maccarone** (Texas Tech U.) <Thomas.Maccarone@ttu.edu>,
**Kristin C. Madsen** (Caltech) <kkm@caltech.edu>,
**Richard Mushotzky** (U. MD) <richard@astro.umd.edu>,
**Angela Olinto** (U. Chicago) <aolinto@uchicago.edu>,
**Peter Plavchan** (George Mason U.) <pplavcha@gmu.edu>,
**Levon Pogosian** (Simon Fraser U.) <levon@sfu.ca>,
**Andrew Ptak** (NASA GSFC) <andrew.ptak@nasa.gov>,
**Paul Ray** (NRL) <paul.ray@nrl.navy.mil>,
**Graca M. Rocha** (JPL) <graca.m.rocha@jpl.nasa.gov>,
**Paul Scowen** (Arizona State U.) <paul.scowen@asu.edu>,
**Sara Seager** (MIT) <seager@mit.edu>,
**Massimo Tinto** (JPL) <massimo.tinto@jpl.nasa.gov>,
**John Tomsick** (UC Berkeley) <jtomsick@berkeley.edu>,
**Gregory Tucker** (Brown) <Gregory_Tucker@brown.edu>,
**Mel Ulmer** (Northwestern) <m-ulmer2@northwestern.edu>,
**Yun Wang** (Caltech/JPL) <wang@ipac.caltech.edu>,
**Edward J. Wollack** (NASA/GSFC) <edward.j.wollack@nasa.gov>

# THE CASE FOR PROBE-CLASS NASA ASTROPHYSICS MISSIONS

**EXECUTIVE SUMMARY:**

Astrophysics spans an enormous range of questions on scales from individual planets to the entire cosmos. To address the richness of 21$^{st}$ century astrophysics requires a corresponding richness of telescopes spanning all bands and all messengers. Much scientific benefit comes from having the multi-wavelength capability available at the same time. Most of these bands, or measurement sensitivities, require space-based missions. Historically, NASA has addressed this need for breadth with a small number of flagship-class missions and a larger number of Explorer missions. While the Explorer program continues to flourish, there is a large gap between Explorers and strategic missions.

A fortunate combination of new astrophysics technologies with new, high capacity, low $/kg to orbit launchers, and new satellite buses allow for cheaper missions with capabilities approaching strategic mission levels. NASA has recognized these developments by calling for "Probe-class" mission ideas for mission studies. Twenty-seven proposals were received and 10 were funded. The submissions spanned most of the electromagnetic spectrum from GeV gamma-rays to the far infrared, and the new messengers of neutrinos and ultra-high energy cosmic rays. The key insight from the Probes exercise is that order-of-magnitude advances in science performance metrics are possible across the board for initial total cost estimates in the range $0.5B - $1B.

**We advocate that the Astro2020 Decadal recommend a new line item for Probes be instituted in the NASA Astrophysics Division budget** within the wedge for large new missions. This recommendation would be in line with the #2 priority of the 2010 Decadal in favor of a vigorous Explorer program. This new Probe-class mission line would set a mission cost cap, as in the successful NASA Planetary Division's New Frontiers and Discovery programs. The Probes line needs to be a significant fraction of the budget over the decade covered by Astro2020. Probes will have costs in the range from just above the MIDEX Explorer cap ($250M, without a ~$50M launch vehicle) up to $1B (total), the nominal lower bound for a strategic mission. A cadence of 2 – 3 probes per decade would be desirable, and possible to integrate with a moderate flagship funding line as well. Like the Explorer line the Probes line would need protection against being raided to pay for cost overruns in flagship missions. There are multiple possible ways to implement a Probes line to reap the maximum advantage for science, but a key recommendation is that a "line" of Probes enables multi-mission development over this decade, and into the future, with flexibility to address new and broad astrophysics topics.

# 1. KEY SCIENCE GOALS AND OBJECTIVES

The breadth of 21$^{st}$ Century astrophysics is staggering. The field is now truly multi-wavelength and multi-messenger. The recent detections of neutrinos from a blazar and gravitational waves from both merging black holes and neutron stars are only the latest surprises that the universe has given up to our increasingly sophisticated instrumentation.

This richness is too great to be confined to a few questions. The full story of how the universe began in a Big Bang and led to stars, galaxies, black holes, planets and, eventually, life is becoming clearer, but with huge unknowns. The nature of the Dark Sector (matter and energy), the origin of seed black holes, the seemingly stubborn barriers to planet formation, and the conditions needed for life to begin are all mysteries.

No single telescope can address this breadth in full. Multiple missions over a range of sizes are needed to achieve balance in the program between fields, in order to maintain astronomy as the vigorous science it has been for the past several decades.

A suite of telescopes is needed. As per the 2017 National Academies report "Powering Science: NASA's Large Strategic Science Missions", flagships have a critical role to play. But given the breadth of scientific return across wavelengths and scientific areas, multiple missions are required, and they cannot all be flagships nor take decades to develop.

For a long time the only formal channel for proposing smaller missions has been the Explorer Program. This program is healthy, with ~4 missions/decade. While growth there would be well-justified given the number of selectable proposals, the Explorer program only supports missions up to $150 M (SMEX) or $250 M (MIDEX) without launch vehicles. The budgets, and launch capabilities (often only to low Earth orbit), fill only a specific niche in the astrophysical discovery space.

The question is whether missions smaller than flagships and larger than Explorers could achieve the ambitious science goals that astronomers seek. Certainly It would be strange if important science can be done for less than $300 M, and important science can be done for more than $1 B, but that in the range $300 M – $1 B is a gap. Instead, as shown by the strong response to the NASA probe call and the quality of the proposals, the answer is a clear YES! Probe-class missions can be both ambitious and are affordable within a broad-based program that provides flexibility for the community. These probes will be making an order-of-magnitude or more gain over their predecessor's measurements – or completely new measurements not possible at smaller scales like Explorers.

A series of probes may yield more great science than a Flagship of the same total cost as:
1. "more bang for the buck": a large observatory with multiple instruments (e.g. Hubble) gets less time per instrument, and each instrument costs more per performance capability due to the enhanced integration and testing needed;
2. lowered risk: if one fails we still have the others;
3. better launch schedule: the first is ready much sooner, and quite plausibly the whole program within the same time frame as the large Flagship mission;

4. lower cost: due to the smaller launch vehicle, and class B requirements rather than class A.
5. better optimization- the telescopes can be optimized for the task rather than being driven by the most severe requirements.

## 2. TECHNICAL OVERVIEW: PROBE-CLASS MISSIONS

NASA Astrophysics has flown previous Probe-class (i.e. ~$1 B) missions, as Paul Hertz has pointed out[1]: COBE, RXTE, Fermi, Kepler (originally selected as a Planetary Division Discovery mission), and Spitzer. All of them were highly productive. Spitzer was one of the "Great Observatories".

But has the potential of this mission class been exhausted? To give a more specific answer to the question, NASA Astrophysics called for proposals to study "Probe-class" missions in the $0.5 B – $1 B (including launch and Phase E) range. Twenty-seven proposals were submitted to NASA[2], and 10 were selected for more detailed study. The reports from these studies are all now available[3]. Several other studies have been funded by NASA that are also in the "probe-class", such as the Solar System Exploration Research Virtual Institute (SSERVI) study of a lunar farside radio array for astrophysics, and two exoplanet studies previously done (Exo-C and Exo-S) looking at dedicated $1B-class exoplanet direct imaging missions.[4]

*Table 1: Completed Probe Studies*

| Probe Study | Band | Closest Predecessor |
|---|---|---|
| AXIS | X-ray | Chandra |
| CDIM | Near-mid-IR | SPHEREx, JWST |
| CETUS | UV | GALEX, HST |
| Earthfinder | Near-IR | Ground-based radial velocity |
| GEP | Mid-IR, Far-IR | Herschel, Spitzer |
| PICO | CMB | Planck |
| POEMMA | Cosmic rays, neutrinos | Auger |
| Starshade | Optical/NIR | WFIRST |
| STROBE-X | X-ray | RXTE, NICER |
| TAP | X-ray, IR, gamma | Swift |
| Farside# | Radio | LWA,MWA , LOFAR, SunRISE |
| Exo-C* | Optical/NIR | WFIRST |
| Exo-S* | Optical/NIR | WFIRST |

# Farside was funded separately from the other 10 probes through a SSERVI award, but is in family with the main probe set. The PI is Jack Burns (U.Colorado).

* These were Probe studies done by ExEP 2 years ago – the Exo-C PI is Karl Stapelfeldt and the Exo-S PI is Sara Seager (MIT) (EXO-S evolved into the Starshade Rendezvous Probe).

---

[1] Presentation at "The Space Astrophysics Landscape in the 2020s, slide 19, URL: https://www.hou.usra.edu/meetings/landscape2019/presentations/Hertz.pdf
[2] URL: https://pcos.gsfc.nasa.gov/physpag/probe/probewp.php and https://cor.gsfc.nasa.gov/copag/probe-study.php
[3] URL: https://science.nasa.gov/astrophysics/2020-decadal-survey-planning
[4] https://exoplanets.nasa.gov/exep/studies/probe-scale-stdt/

The key insight from these Probe studies is that order-of-magnitude advances in science performance metrics are possible across the board for initial cost estimates in the range $0.5B - $1B. This is possible because of investments in new instrument technologies and leveraging commercial satellite buses allows for missions with capabilities approaching flagship mission levels, but at a significantly lower cost. The advent of new, high-capacity, low $/kg-to-orbit launchers from SpaceX, ULA, and Blue Origin, will continue to bring down the cost of these capable missions by encouraging rideshare opportunities for multiple assets on a single launch, or affording the flexibility to optimize design and cost with more relaxed launch mass and fairing constraints.

The probe-class mission studies are credible examples that demonstrate the wide variety of possible Probe-class missions[5], but by no means exhaust the possibilities for scientific discovery at an affordable price point, while achieving programmatic balance. The Probe studies showed significant scientific resilience as well, allowing the scope and design to be optimized for a given cost target or launch constraint.

There is a perception in parts of the community that Probe science is necessarily highly targeted whereas flagships have broad science. While almost all the studied Probes derive their parameters from a small number of key science objectives, the resulting science impacts many areas of astrophysics in virtually every case, and serves a broad community of astrophysicists.

SUMMARY OF COMPLETED PROBE STUDIES

ADVANCED X-RAY IMAGING SATELLITE (AXIS) is a major improvement over *Chandra* — with higher-resolution imaging over a larger field of view at much higher sensitivity, and agile operations allowing *Swift*-like transient science. Science includes: growth and fueling of supermassive black holes; galaxy formation and evolution; microphysics of cosmic plasmas.

COSMIC DAWN INTENSITY MAPPER (CDIM) will transform our understanding of the era of reionization when the first stars and galaxies formed, and UV photons ionized the neutral medium. CDIM uses wide area spectro-imaging surveys to provide redshifts of galaxies and quasars during reionization and crucial information on physical properties.

COSMIC EVOLUTION THROUGH UV SPECTROSCOPY (CETUS) is a 1.5-m wide-field UV telescope that will be a worthy successor to Hubble. With its wide-field camera, multi-object spectrograph, and high-resolution echelle spectrograph, CETUS will maintain observational access to the ultraviolet (UV) after Hubble and also provide new and improved capabilities.

EARTHFINDER will perform high precision (cm/s) radial velocity (PRV) measurements by taking advantage of: broad wavelength coverage from the UV to NIR; extremely compact, highly stable and efficient spectrometers; laser-based wavelength standards; high cadence observing to minimize sampling-induced aliases; and absolute flux stability for line-by-line analysis.

---

[5] Taken from the probe study reports directly (Executive summary or Introduction).

GALAXY EVOLUTION PROBE (GEP) will use the mid and far-IR to map the history of galaxy growth by star formation and accretion by super-massive black holes and their inter-relation, and will measure the evolution of the interstellar medium and build-up of life-enabling elements over cosmic time.

PROBE OF INFLATION AND COSMIC ORIGINS (PICO) is an imaging polarimeter that will scan the sky for 5 years in 21 frequency bands spread between 21 and 799 GHz. It will produce full-sky surveys of intensity and polarization with a final combined-map noise level equivalent to 3300 Planck missions for the baseline required specifications, performing as 6400 Planck missions.

PROBE OF EXTREME MULTI-MESSENGER ASTROPHYSICS (POEMMA) observes the Earth's atmosphere to see extensive air showers from cosmic rays >20 EeV and cosmic neutrinos >20 PeV to study the origin of the highest-energy particles; neutrino emission of extreme transients; particle interactions at extreme energies; luminous transient events; and exotic particles.

STARSHADE RENDEZVOUS PROBE, operated in formation with the WFIRST observatory can perform space-based direct imaging capable of discovering and characterizing exoplanets around our nearest neighbor star systems. This first-of-its-kind combined mission will enable a deep-dive exoplanet investigation around these neighbor star systems

SPECTROSCOPIC TIME-RESOLVING OBSERVATORY FOR BROADBAND ENERGY X-RAYS (STROBE-X) combines huge collecting area, broad energy coverage, high spectral & temporal resolution, & agility to measure mass and spin & map accretion for all black hole masses; map the neutron star mass-radius relation; identify & study X-ray counterparts of multiwavelength & multi-messenger transients.

TRANSIENT ASTROPHYSICS PROBE (TAP) will characterize electromagnetic counterparts to Gravitational Waves for mass scales from neutron stars to $10^9$ $M_\odot$ Supermassive Black Hole Binaries, and many time-domain astrophysical phenomena. TAP is an agile multi-instrument platform with wide-field X-ray detectors, $4\pi$ gamma-ray monitors, as well as X-ray and wide-field IR telescopes.

FARSIDE leverages the Lunar Gateway infrastructure to enable a low frequency 128-node radio array, completely deployed robotically, for studies of magnetic fields in known exoplanetary systems.

EXO-C would use a dedicated 1.4 m telescope and coronograph to spectrally characterize at least a dozen RV planets, search >100 nearby stars at multiple epochs for planets down to $\sim 3\times 10^{-10}$ contrast, characterize mini-Neptunes, search the $\alpha$ Centauri system, and image hundreds of circumstellar disks.

EXO-S would use a dedicated 1.1 m telescope and 30 m starshade for direct imaging and spectral characterization, of giant planets down to Earth-size, and study complete planetary system as well as circumstellar dust. The Starshade Rendezvous Probe concept originally came from this study and was studied in more detail as per above.

There were about two dozen Probe mission white papers submitted that were not selected. While some of them may well not have been selectable, and some were duplicative, a number clearly were selectable and were not chosen for want of program funding. Table 2 lists these missions from on the Cosmic Origins and PCOS web sites (see URLs above). The variety of mission concepts indicates that there is great potential depth to the Probe-class mission class.

*Table 2: A selection of Probe White Papers submitted to the NASA call for concept studies,(based on public web sites), demonstrate the great interest and broad diversity of science possible in this intermediate mission class.*

| Mission | Band | PI |
|---|---|---|
| Death of Massive Stars (DoMaS) | Gamma, X-ray, IR | Pete Roming |
| Inflation Probe | mm, sub-mm | Ed Wollack |
| X-ray Grating Spectroscopy Probe | X-ray | Randall McEntaffer |
| HEX-P | Hard X-ray | Fiona Harrison |
| mHz Gravitational waves | GW | Massimo Tinto |
| 99 Luftballons | Near-IR | Tim Eifler |
| Advanced Particle Telescope | Cosmic rays | James Buckley |
| Time-domain Spectroscopic Observatory | X-rays, IR, gamma | Josh Grindlay |
| Wide-field X-ray Probe | X-rays | Andy Ptak |
| AMEGO | MeV, GeV | Julie McEnery |
| Probe-class Gravitational Wave Observatory | GW | Sean McWilliams |
| GreatOWL | Cosmic rays | John Mitchell |
| Probe-class Far-IR | Far-IR | C. Bradford |
| Dark Ages and Cosmic Dawn | Low ν radio | Joe Lazio |
| WFXIS | X-ray | Mel Ulmer |
| ALLEGRO | X-ray | Mel Ulmer |
| ForEST | Optical/near-IR | Howard MacEwen |
| HORUS | UV/optical | Paul Scowen |
| Deep Survey Telescope | Near-IR | Fred Hearty |
| SHARP-IR | Far-IR | S. Rinehart |
| ORION | UV/optical | Paul Scowen |
| ATLAS | Near-IR, Mid-IR | Yun Wang |
| Cosmic Origins & Destiny | Far-IR | Christopher Walker |
| NG-SUVO | UV/optical | Mel Ulmer |
| SPECTRAS | Mm, sub-mm | Paul Goldsmith |

**A BALANCED PROGRAM FOR THE 2020'S**

The current bifurcation of NASA Astrophysics missions between a very limited number of multi-billion dollar flagships and the roughly 10-50 times cheaper Explorers has led to an unbalanced program. There are alternatives. The NASA Planetary Division has the cost-capped $1 B New Frontiers[6] and $500 M Discovery programs[7], in addition to the program's flagship missions. The New Frontiers missions to date are Juno, New Horizons, and OSIRIS-Rex, and there have been many Discovery missions. Cost caps do not typically include the launch vehicle, or Phase E operations, but the astrophysics Probe studies did include these (substantial) items.

Given the example programmatics from NASA's Planetary Division, the wealth of strong Probe proposals, and the feasibility demonstrated by the selected studies there would seem to be no technical obstacle to adopting a similar approach, with minor adjustments, for the Astrophysics program.

Astrophysically, to maintain a broad-based program, a cadence of 2 - 3 Probe-class missions per decade would be both plausible and desirable. In order to encourage a range of Probe cost levels, a division into cost sub-categories analogous to the New Frontiers/Discovery (or MIDEX/SMEX) division could be implemented.

**ADVANTAGES OF PROBES**

These medium-sized missions spread throughout the decade has a number of advantages that will help to ensure U.S. leadership in astrophysical science:

1. <u>scientific</u> (responding to emerging science areas, science breadth through diversity, opportunity for vast GO programs);
2. <u>participatory</u> (multiple institutions and industries would be engaged across a spectrum of capabilities);
3. <u>financial</u> (smoothing funding profiles across the decade through diversity of timelines and peak spending years, lower cost missions have lower cost risk typically);
4. <u>a deep bench</u>. An increased number of US scientists with experience in proposing for and successfully managing large proposals. Increasing the 'bench' of investigators who can PI scientific missions will increase the diversity of scientific ideas for all mission classes, resulting in a more innovative scientific program across the board;
5. <u>buying down risk</u>. More ambitious, possibly flagship, versions will have some of the risk retired;
6. <u>enables technology development</u>. Development across multiple bands more continuously without hiatus.

**IMPLEMENTATION OPTIONS**

There are multiple possible ways that the Decadal might recommend implementing a Probes line to reap the maximum advantage for science. If the Astro2020 Decadal Survey recommends the creation of a Probe line, there is the opportunity to advise NASA on how to implement such a program as well. Several options have been discussed:

1. Follow the example of the New Frontiers program for a new Astrophysics Probes program, with competed, PI-led mission concepts that are constrained to a subset of science priorities set by the most recent Decadal Survey. This allows the Decadal to guide a few areas of high priority discovery space that could credibly be achieved in this price point.

2. The Decadal Survey could, in theory, recommend specific Probe missions for implementation as part of a balanced, strategic program, avoiding a direct competitive AO. NASA could then assign them to an implementing NASA center, and compete instruments and/or science teams. Two cautions apply to this approach: (a) Probes may more likely to be treated as Class A missions, leading to cost overrun issues; and (b) the science team will be less involved in mission details, leading to poorer communications between the engineers and the scientists.

3. Emulate the Explorer and Discovery programs for which there are no restrictions (or prioritization from the Decadal) on science, which allows the most flexibility. There could be concerns about the potential burden on the proposing community for an open call on such large missions, but it seems likely that the nature of ~$1B missions will limit credible proposers to a manageable levels for community. A concern is that many proposals may be submitted at considerable effort for each. A 2-phase submission process might relieve some of the pressure.

---

[6] URL: https://science.nasa.gov/solar-system/programs/new-frontiers
[7] https://science.nasa.gov/solar-system/programs/new-frontiers; https://science.nasa.gov/solar-system/programs/discovery

No single approach need be chosen. It may be that the Decadal concludes that a more guided program is appropriate for the near-term, but that later Probe missions should be unconstrained in order to be able to respond to the changing astrophysics landscape over the latter part of the study.

A more detailed analysis of programmatic options for creating a Probe line is warranted in order to maximize scientific return and insure programmatic balance.

**4. TECHNOLOGY DRIVERS**

Missions in the Probe-class, as evidenced by the completed study reports, do rely on advanced technologies. Rather than the current situation, where competed Explorers eschew new technology as much as possible, and large flagships take all the technology cost risk, Probes could allow a balanced approach to technology.

To provide probes with access to advanced technology, the SAT program (or a related program) will need augmented funding. Scaling from flagships, something like 3 - 5% of the probe lifecycle cost would seem appropriate. At 2 probes per decade, this would require an additional ~$5M/year in Astrophysics SAT funding. This is not negligible, but not impossible. E.g. the APRA program just received a $5M/year increase to support cubesats.

Advancing technology readiness across such a broad spectrum of concepts would be balanced by direct NASA funding through something like the Strategic Astrophysics Technology (SAT) program. If science areas are driven by Decadal priorities, it would enable more strategic investing by NASA. But the nature of a competition would still incentivize proposers to invest in their own technologies and avoid very high-risk technologies that could blow a missions cost cap. Encouraging technological breakthroughs where they make the most impact in a price point that allows teams to actually manage advancement is a great opportunity for Probes.

## 5. ORGANIZATIONS, PARTNERSHIPS, AND CURRENT STATUS

As evidenced by the breadth of university, government, and industrial partners in the multiple probe studies completed and submitted to date, it is clear that a large, diverse collection of the community would benefit from a series of Probe missions.

In addition, while secondary to scientific strength, robust community participation in the mission science is desirable, either through a GO observing program where appropriate or through a significant early archival research program. Most Probes are amenable to one or both forms of such a program (either observationally, like Spitzer, archivally, like Fermi, or with data product releases, like Gaia). Existing data centers like STScI, the CXC, and IPAC could disseminate Probe data the community broadly, as per NASA guidelines. Funding to support probe science by GOs would be needed.

## 5. COST AND SCHEDULE

The wedge anticipated for large new mission development is ultimately ~$7 B over a decade. Initial startup may be delayed in the 2020s due to already selected missions. The ten Probe studies included within their total cost cap of $1B both of launch and of Phase E operations. Hence the suggested cadence of 2 - 3 Probe-class missions per decade appears plausible at ~$2B - $2.5B. This would require a range of Probe cost levels to average to ~$0.8B to launch. A division of probes into cost sub-categories analogous to the New Frontiers/Discovery or MIDEX/SMEX division could be implemented to encourage a range of mission costs. A Probe line on this scale would still allow the development of a ~$5B Flagship mission in the same decade, subject to funding peak compatibility.

The concurrency gained from having multiple powerful observatories operating together is a substantial gain over having them sequentially. A cadence of 3 Probes plus one Flagship per decade going forward would, assuming extended missions, produce a revival or continuation of the breadth of capability provided by the Great Observatories.

The Probe-class mission line would need to be protected against being raided to pay for cost overruns in flagship missions, in a manner similar to the Explorer program. Probe-class missions spread throughout the decade would smooth funding profiles across the decade through the diversity of their timelines and spending peaks. Also lower cost missions often have lower risk.

## 6. CONCLUSIONS

The key question for Probes is whether there is compelling science in the wide cost range between Explorers and Flagships. It seems self-evident that there is no scientific desert between those extremes. The NASA Probe studies, summarized here and submitted separately, provide clear examples that yes, there is a rich diversity of forefront science doable at the Probe scale. Some of the science can only be done efficiently at that scale, driven by the scientific requirements for discovery and understanding. Finally these studies demonstrate that for a given cost or technical target, the astrophysics community has the creativity to meet that target. Whatever the cost cap the scientists and engineers in our community will meet the challenge for Probes and provide a compelling, continuing program for astrophysical discovery.